\documentclass[12pt,preprint]{aastex}
\usepackage{threeparttable}

\begin{document}

\title{The intriguing giant bow shocks near HH 131
\footnote{Based in part on data collected at Subaru Telescope,
which is operated by the National Astronomical Observatory of
Japan.}}

\author{Min Wang\altaffilmark{1,2},Junichi Noumaru\altaffilmark{3}, Hongchi Wang\altaffilmark{1,2}, Ji Yang\altaffilmark{1,2},
and Jiansheng Chen\altaffilmark{2,4,5}}

\altaffiltext{1}{Purple Mountain Observatory,
 Academia Sinica, Nanjing 210008, China; mwang@pmo.ac.cn,
  hcwang@pmo.ac.cn, jiyang@pmo.ac.cn}
\altaffiltext{2}{National Astronomical Observatories, Chinese
Academy of Sciences, A20 Datun Road, Chaoyang District, Beijing
100012, China}
\altaffiltext{3}{SUBARU project, Japan; noumaru@subaru.naoj.org}
\altaffiltext{4}{BAC: Chinese Academy of Sciences - Peking
University joint Beijing Astrophysical Center, Beijing 100871,
China}
\altaffiltext{5}{Astronomy Department, Peking University, Beijing
100871, China}

\begin{abstract}
Using the High Dispersion Spectrograph (HDS) at the Subaru
Telescope, echelle spectra of two giant arcs, i.e. nebulosities Cw
and L (hereafter Nebu.~Cw and L, or simply Cw and L) associated
with HH 131 in Orion are presented. Typical emission lines of
Herbig-Haro (HH) objects have been detected towards Nebu.~Cw with
the broadband filter KV~408. With the low dispersion spectrograph
at the National Astronomical Observatories (NAO) 2.16~m telescope,
spectra of Nebu.~C, L and K are obtained, which also show strong
[S~$\rm{II}$]$\lambda$$\lambda$ 6717/6731, H$\alpha$ and
[N~$\rm{II}$]6583 emission lines. Position-velocity distributions
of Cw and L are analyzed from the long-slit spectra observed with
the HDS H$\alpha$ narrowband filter. The fastest radial velocity
of Cw is $V_r\sim$--18.0~km~s$^{-1}$. When the flow at L goes to
the south, it slows down. The fastest radial velocity of L has
been observed of --45.0~km~s$^{-1}$ and the slowest value is about
--18.3~km~s$^{-1}$, the radial velocity gradient is about
200~km~s$^{-1}$~pc$^{-1}$. The similarity of the fastest radial
velocity of Cw to the slowest value of L and their positional
connection indicate that they are physically associated. There is
a tendency for the entire flow to become less excited and less
ionized when going further to the south (i.e., from Nebu.~K, L to
C), where the most extended (and presumably evolved) objects are
seen. The electron densities of all the observed nebulosities are
low ($n_{\rm e}\sim$10$^2$~cm$^{-3}$). Double kinematic signatures
have been found in Cw from its [N~$\rm{II}$]6583 profiles while
the observed H$\alpha$ profiles of Cw are almost symmetric. Bow
shock models appear to agree with the observed position-velocity
diagrams of the [N~$\rm{II}$] spectra better than H$\alpha$
spectra, and a bow shock with its wing, apex and postshock has
been possibly revealed near Cw from the [N~$\rm{II}$] emission.
With the suggestion that these arcs are HH shocks possibly ejected
out of the Orion~A molecular cloud by an uncertain source, their
spectra show low to intermediate excitation from their diagnostic
line ratios.
\end{abstract}


\section{Introduction}
Mass outflows emerge from young stellar objects (YSOs) at very
early stages of star formation processes. Mass outflows are traced
by Herbig-Haro (HH) objects at optical wavelengths, while
molecular hydrogen and [Fe~$\rm{II}$] emission are seen at near
infrared wavelengths. At mm-wavelengths bipolar molecular outflows
are best observed in the lines of CO. HH objects are collisionally
excited nebulae produced by outflows ejected by YSOs, and they
trace either shocks where outflows ram the quiescent interstellar
medium (terminal working surfaces) or shocks produced by colliding
fluid elements ejected from the source at different velocities and
times (internal working surfaces) (Raga et al. 1990). The
conception of parsec scale HH flows has altered our understanding
of optical jets and HH objects since Bally \& Devine firstly
discovered the giant HH~34 flow which extends about 3.4~pc (Bally
\& Devine 1994; Reipurth et al. 1997; Devine 1997). The total
extension of an HH flow is much larger than originally thought
(Mundt \& Fried 1983), surpassing 1~pc in most cases. Thanks to
the technical development of larger CCD arrays, wide field surveys
of optical outflows have been recently carried out covering dozens
of square degrees in nearby star forming regions (Yan et al. 1998;
Zhao et al. 1999; Yang \& Yao 2000; Wang et al. 2000, 2001; Wu et
al. 2002; Sun et al. 2003; Walawender et al. 2005). Spectrographic
measurements and analyses in terms of bow shock models provide a
comprehensive picture of the outflow phenomenon of YSOs (Schwartz
1975; Hartigan et al. 1987; Raga et al. 1996). However, it is not
yet clear in how far the pc scale HH objects impact on the
surrounding interstellar medium (ISM). HH~131 and its nearby giant
bow shocks possibly provide the most intriguing targets, which are
located well outside the Lynds 1641 (hereafter L1641) dark cloud,
i.e., about 1.5$^\circ$ southwest of L1641. In L1641 not only the
largest number of HH objects (in all 70 from the list by Reipurth
1999), and the prototypical HH objects, HH~1 and HH~2, but also
the first pc scale HH flow, HH~34, have been discovered. All of
these HH objects are situated inside or very close to the boundary
of the Orion~A giant molecular cloud (GMC). With narrowband
imaging Ogura (Ogura 1991) discovered HH~131 and some large
nebulosities like A, B, C, ..., K and L, and suggested that those
nebulosities might be physically HH objects. The linear scale of a
single bow shock of them is even larger than 3~pc (we adopt a
distance of 450~pc). Intending to study their nature, with the
SUBARU telescope we have made spectrographic observations toward
Cw and L. Cw is located slightly west to the given position of C
(see the image shown in Fig.~1, which was derived with the
0.6/0.9~cm Schmidt telescope at Xinglong Station of National
Astronomical Observatories (NAO), Chinese Academy of Sciences).
Low-dispersion spectra of C, L and K were also obtained with the
spectrograph mounted at the NAO 2.16~m telescope.

\section{Observation and Data Reduction}

With the High Dispersion Spectrograph (HDS) at the Subaru
Telescope, the observations of Nebu.~Cw and L were carried out on
the night of October 22/23 2001. The filters used in the
observations were a broadband one named StdYc or KV~408, and a
narrowband one named StdHa or H$\alpha$. Table 1 lists the
observing targets and instrumental parameters. The slit with a
width of 1.0$''$ provided a spectral resolution of 36000
corresponding to a velocity resolution of 8.3~km~s$^{-1}$ that was
confirmed by sky lines. The observed spectra were wavelength
calibrated using exposures to a Th-Ar comparison lamp. The
accuracy of wavelength calibration is better than 0.01~\AA, or
0.5~km~s$^{-1}$ at H$\alpha$.

We obtained spectra for Cw with the KV~408 filter, the slit length
was set as 5$''$, its exposure time was 1800 sec. Flux calibration
was achieved using echelle spectra of the standard star Hz~15
(Stone 1977), and its exposure time was 120 sec. The wavelength
range covered 4390--5710~\AA\, and 5810--7130~\AA\, for CCD1 and
CCD2, respectively. The linear dispersion of the CCD near
6600~\AA\, was $\sim$0.04~\AA~pixel$^{-1}$ or 1.8~km~s$^{-1}$.

Narrowband long-slit spectra were taken for Cw and L (see Fig.~1)
with the slit at position angles of 0$^\circ$ and 150$^\circ$,
respectively. The slit length was 60$''$. The StdHa/H$\alpha$
filter was used, which was centered on 6580~\AA~ with a passband
of 54.6~\AA, covering H$\alpha$ and
[N~$\rm{II}$]$\lambda$$\lambda$6548/6583 emission lines. The
integration time of each position was 2$\times$0.5 hours. The
angular scale on the CCD frame was $\sim$0.27$''$~pixel$^{-1}$
along the direction of the slit (CCD 2$\times$2 binning mode
used).

Except of overscanning dealt with a procedure provided by the
SUBARU project group, data reduction has been made with the IRAF
package, including bias-subtraction, flat-fielding, spectra
extraction, wavelength and flux calibrations, and velocity
corrections. For the observations with the H$\alpha$ filter,
wavelength calibration has to be made according to two echelle
orders (Order 91 and 90) separately instead of most likely just
for Order 91. Then we co-add these two calibrated spectra into one
spectrum (priv. comm. with Aoki Wako and Akito Tajitsu).

In data reduction we noticed the problem caused by contamination
of sky line emission. Because of the limited slit length used for
the broadband mode, subtraction of the sky emission spectrum was
impossible since the observed object fills the entire slit
aperture. In the case of the narrowband mode, the sky emission was
possibly subtracted, because the relatively much longer slit
aperture used was covering regions where the signal from the
object was much weaker than that at peak. These regions were taken
for establishing the `sky spectrum' (including a small
contribution from the extended object), which was then subtracted
from the `object spectrum'. As a consequence, the sky emission
could be slightly over- or under-subtracted (see little tips or
valleys at $\sim$ 20~km~s$^{-1}$ in Fig.~7 and Fig.~8).

Low-dispersion spectroscopic observations of Nebu.~C, L and K were
made with the NAO 2.16~m reflector using a Cassegrain spectrograph
during 1999 December 5--9. The grating used was 100~\AA~mm$^{-1}$.
The slit was 4.0$'$ long with a width of 2.0$''$. The resulting
spectral resolution is $\sim$7\AA. All slit orientations followed
 the north-south direction.

\section{Results and Discussions}
\subsection{Emission lines}
\subsubsection{Low-dispersion spectra}
Fig.~2 displays the low resolution spectra of Nebu.~C, L and K,
which were observed with the NAO 2.16~m telescope. Line
identifications and integrated line
intensities are presented in Table~2 .
The [S~$\rm{II}$]$\lambda$$\lambda$ 6717/6731, H$\alpha$ and
[N~$\rm{II}$]6583 lines have been clearly detected and show strong
emission towards all the three nebulosities. Although the
detections of [N~$\rm{II}$]6548 and H$\beta$ are certain, there
are large uncertainties in their line intensities than in the
former four lines. The detection of [O~$\rm{I}$]6300 emission from
L is tentative due to strong contamination of sky light. Continuum
emission was not detected from any of the three nebulosities. The
clearly detected emission lines toward C, L and K are
characteristic signatures of HH objects with strong forbidden
lines and no continuum emission. No reddening correction has been
applied to the emission line intensities through the entire paper.
Such a correction is not critical in the interpretation of line
ratios which are close in wavelength, e.g., those of H$\alpha$ to
[S~$\rm{II}$] or [S~$\rm{II}$]$\lambda$6717/$\lambda$6731.

The electron densities ($n_{\rm e}$) listed in Table~2 have been
deduced from the [S~$\rm{II}$]$\lambda$6717/$\lambda$6731 line
intensity ratios assuming an electron temperature of 10,000~K
(Osterbrock 1989). They tend to decrease from about 400 to less
than 50 cm$^{-3}$ for K, L and C.

\subsubsection{HDS spectra}

Towards Cw, the [S~$\rm{II}$]$\lambda$$\lambda$ 6717/6731,
H$\alpha$, [N~$\rm{II}$]$\lambda$$\lambda$ 6548/6583,
[O~$\rm{I}$]$\lambda$$\lambda$ 6300/6363 and H$\beta$ emission
lines have been detected in our HDS echelle spectra (Fig.~3).
Continuum emission was not detected. The H$\alpha$ and H$\beta$
lines present broader profiles than those of [S~$\rm{II}$] and
[N~$\rm{II}$] while the case of the [O~$\rm{I}$]6300/6363 lines is
less clear. Gaussian fitting parameters are given in Table~3. The
detected lines suggest that Nebu.~Cw possesses quite a complex
kinematic structure. At least two velocity components are seen at
about --15~km~s$^{-1}$ (It is a heliocentric velocity when not
specially noted through the entire paper) and 0~km~s$^{-1}$,
respectively. There is likely also a third component around
20--27~km~s$^{-1}$, which is most likely from the sky emission, or
partly from the Orion star forming region. So the giant bow shock
Cw is slightly moving toward us with the extreme velocity of
--17.5~km~s$^{-1}$ seen in the [N~$\rm{II}$]6583 emission line.
The complex velocity structure of Cw will be discussed in detail
in Sect.~3.3.

The [S~$\rm{II}$], [N~$\rm{II}$] and [O~$\rm{I}$] lines have all
emission peaks at about --15~km~s$^{-1}$ (in Table~4). From the
line emission peak intensities, we estimate that:
[S~$\rm{II}$]$\lambda\lambda$(6717+6731)/H$\alpha\sim$1.55 and
[N~$\rm{II}$]6583/[S~$\rm{II}$]$\lambda\lambda$(6717+6731)$\sim$0.41.
According to the integrated fluxes (in Table~3),
[S~$\rm{II}$]$\lambda\lambda$(6717+6731)/H$\alpha\sim$1.54 (1.33
without the third velocity component included, and the third
component of H$\alpha$ is never counted for ratio calculations
because of the certainty of H$\alpha$ emission from the sky) and
[N~$\rm{II}$]6583/[S~$\rm{II}$]$\lambda\lambda$(6717+6731)
$\sim$0.38 (0.31).

According to the total fluxes,
[S~$\rm{II}$]$\lambda$6717/$\lambda$6731$\sim$1.34 (1.37). From
the peak intensities in Table 4,
[S~$\rm{II}$]$\lambda$6717/$\lambda$6731$\sim$1.26. This implies
that the electron density is about 80--170~cm$^{-3}$. The
non-detection of [O~$\rm{III}$]~5007 implies that the shock
velocity is less than 100~km~s$^{-1}$ (Hartigan~et~al. 1987).

\subsection{What is the nature of the nebulosities?}

In order to explore the true nature of these nebulosities we have
plotted the line ratios of H$\alpha$/[S~$\rm{II}$](6717+6731)
versus the ratios of [S~$\rm{II}$]$\lambda$6717/$\lambda$6731 for
Nebu.~C, Cw, L and K in Fig.~4, comparing them with those of HH
objects, compiled by Raga~et~al. (1996), and planetary nebulae
(hereafter PNe), supernova remnants (hereafter SNRs) and
H~$\rm{II}$ regions (adopted from the figures by
Sabbadin~et~al.~1977 and Meaburn \& White 1982). Similar to the
morphology displayed in Fig.~1, in Fig.~4 Nebu.~Cw~(C), K and L
present a physical nature very close to that of SNRs  and far from
that of H~$\rm{II}$ regions or PNe. The strong
[S~$\rm{II}$]$\lambda\lambda$~6717/6731 and the non-detection of
[O~$\rm{III}$]5007 suggest that Nebu.~Cw~(C), K and L cannot be
SNRs (Fesen~et~al. 1985; Fesen \& Hurford 1996). Therefore we
suggest that these nebulosities are HH shocks. In Fig.~4
Nebu.~Cw~(C), K and L have line ratios similar to those of
HH~125~I--K, HH~131, HH~128, HH~111~L, HH~235 (or GGD~35), HH~34MD
and HH~124~A--C. It is especially interesting that HH~125 and
HH~124 also present large bow shocks (Walsh~et~al. 1992). HH~128,
HH~34MD, HH~111~L and possibly HH~235 are parts of individual pc
scale HH objects (Reipurth~et~al. 1997; Ray~et~al. 1990,). One
giant part of the HH~111 flow, i.e. HH~311, extends out to 5.5$'$
(0.7~pc at a distance of 450~pc, Reipurth~et~at. 1997), which has
been previously thought as the largest extension of an HH bow
shock. Nebu.~C extends more than 26~$'$, i.e. 3.4~pc with D$\sim$
450~pc. It is much larger than the former one.

Raga~et~al. (1996) made statistics on a large fraction of optical
spectra of HH objects and derived a quantitative criterion to
divide HH spectra into the high, intermediate and low excitation
categories. A dashed vertical line in Fig.~4 shows the division of
high/intermediate excitation (to the right side) and low
excitation of HH spectra (to the left side) according to the line
ratio of [S~$\rm{II}$](6717+6731)/H$\alpha$ with a value of 1.5.
The intensity ratios of [SII](6717+6731)/H$\alpha$ are 1.6, 1.4,
and 0.9 for Nebu.~C~(Cw), L and K, respectively. With the clear
criterion and the non-detection of [O~$\rm{III}$]5007, their
spectra show low to intermediate excitation. The excitation tends
to decrease from Nebu.~K, L to C, that is, there is a tendency for
the flow to become less excited when going further to the south.

\subsection{Position-velocity distributions of Nebu.~Cw and L}

Figs.~5 and 6 display the position-velocity diagrams obtained from
the long-slit spectra of Nebu.~Cw and L, respectively. We have
detected three emission lines, H$\alpha$ at 6563 and [N~$\rm{II}$]
at 6583 and 6548~\AA. [N~$\rm{II}$]6548 is the faintest of the
three lines, only clearly seen at the brightest positions with the
relative position (hereafter denoted as Y) at --6--0$''$ for both
Cw and L, and is thus not displayed or further analyzed. For each
line one component ($V_3$) at 20--25~km~s$^{-1}$ is relatively
homogeneous through the entire slit and has been identified as sky
emission (Fig.~5 and 6). Similar velocity values of
[N~$\rm{II}$]$\lambda$6548 and [S~$\rm{II}$]$\lambda\lambda$
6717/6731 have also been estimated in Table~3.
In Fig.~5 the other two velocity components are clearly seen in
[N~$\rm{II}$] towards Cw. One is at about 0~km~s$^{-1}$ ($V_2$)
and the other is around --15~km~s$^{-1}$ ($V_1$). The $V_2$
component appears to be blended with some background emission,
which possibly represents emission from the local star forming
region. The $V_1$ component is also likely contaminated with a
fainter similar background. At H$\alpha$ a bright feature has been
observed at the same position as the [N~$\rm{II}$] line with Y
around 0$''$. However, the two velocity components have not been
discerned at H$\alpha$ while the background is strong. In Fig.~6
it is shown that the position-velocity diagrams of both H$\alpha$
and [N~$\rm{II}$] of L are different from Cw. They present less
complex velocity features showing that the emission is possibly
from several spatially separate parts. Dividing each line into
twenty spatially equal parts (each part corresponds to 3$''$),
spectra have been extracted. And Figs.~7 and 8 give their profiles
with the sky emission component ($V_3$) subtracted in the way
described in Sect.~2.

In general, with Declination ($\delta$) increasing, i.e. with Y
increasing in Cw and decreasing in L, both Cw and L tend to have
more negative velocity values (see Fig.~5--8). In the following we
present and discuss their individual features in detail.

At Y=0.0$''$ Nebu.~Cw is brightest with the velocity of
$V_1$~$\sim$--18.0~km~s$^{-1}$. The emission starts to be seen at
Y=--21.0$''$ with $V_2$$\sim+$3.7~km~s$^{-1}$. The $V_2$ component
becomes brighter with increasing Y when Y=--21$''$ --~--15.0$''$.
At Y=--12.0$''$ the profile shows clearly double peaked feature
with $V_2$=--0.1~km~s$^{-1}$ and $V_1$=--17.4~km~s$^{-1}$. Above
Y=--9.0$''$, the $V_1$ component begins to be brighter than $V_2$
while $V_2$ gets closer and closer to $V_1$. At
Y=--6.0$''$--0.0$''$ the $V_1$ emission becomes brightest while
the $V_2$ component disappears; then the $V_1$ emission gets
fainter with Y increasing till it almost disappears at
Y$\sim$15.0$''$. H$\alpha$ is obviously broader than
[N~$\rm{II}$]$\lambda$6583, and likely the corresponding $V_1$ and
$V_2$ components of H$\alpha$ are blended, therefore, the above
variations of velocities and intensities with different positions
are only barely seen. Fig.~9 displays the Gaussian fitting
parameters of the [N~$\rm{II}$]6583 profiles of Cw varying with
positions, including the central velocity in (a), FWHM in (b), the
relative central intensity in (c) and the peak intensity in (d)
where the $V_1$ component is indicated with filled circles and the
$V_2$ one with open symbols.

For Nebu.~L the velocity distribution is different from Cw. Except
of a slight wiggle around Y=6$''$, as a whole with Y increasing
the velocity decreases while the emission intensity has maxima at
three positions (--33.0$''$, --6.0$''$ and 21.0$''$) (see Fig.~6
and 8). At Y=--33.0$''$, Nebu.~L has its fastest velocity value of
--43.4~km~s$^{-1}$ measured in [N~$\rm{II}$]6583, and
--44.5~km~s$^{-1}$ in H$\alpha$. The slowest velocity
$V\sim$--18.3~km~s$^{-1}$ is reached at Y=21.0$''$, which
approximates to $V_1$ of Cw. The fastest radial velocity of Cw is
similar to slowest value of Nebu.~L. Together with their
positional connection, this indicates that they are physically
associated. So when Nebu.~L goes further to the south, it slows
down with a radial velocity gradient of 200~km~s$^{-1}$~pc$^{-1}$,
which is comparable to that of the blue or the red lobe of the
HH~34 flow (Devine et al. 1997).

\subsection{Comparison of observations with bow shock models}
Line profiles and position-velocity diagrams have been
theoretically predicted by Hartigan et al. (1987) and Raga \&
B\"ohm (1986) for a bow shock. They assume that a large fraction
of the bow shock falls in the slit.  In addition, they have showed
that as the slit width changes the predicted line profiles alter
radically, and profiles can also be significantly different from
one line to the other with the same slit width (Figs.~3e, 3t and
3v of Hartigan et al. 1987). For our case, the slits actually
sample a very small region of the enormous arcs. So, when we
compare our observation results with their predictions, it might
be reasonable that some disagreements will occur.

Hartigan et al. (1987) presented theoretical line profiles of bow
shock models with different shock velocities ($V_s$) and
orientations (denoted by $\phi$, which is the angle of the plane
of the sky w.r.t. the axis of symmetry of the bow shock). For
Nebu.~Cw a low shock velocity can be derived from the
[N~$\rm{II}$]6583 line rather than from H$\alpha$ (shock velocity
equals $\Delta{\it V}_{\rm FWZI}\sim$~45.0~km~s$^{-1}$, where
$\Delta{\it V}_{\rm FWZI}$ is the full width at zero intensity
level). Double-peak H$\alpha$ profiles occur only when
$V_s>$150~km~s$^{-1}$ and $\phi>$45$^\circ$ ($\phi$ is adjusted to
follow the definition by Raga \& B\"ohm (1986)). However, such
double kinematic signatures have been found in the
[N~$\rm{II}$]6583 profiles of Cw. The observed H$\alpha$ profiles
of Cw are almost symmetric. Comparing the computed
position-velocity diagrams of [N~$\rm{II}$]6583 with those of
H$\alpha$ predicted by Raga \& B\"ohm (1986) in their Figs.~2 and
3, we note that at each $\phi$ the [N~$\rm{II}$]6583 diagram has
more delicate structures at peaks than H$\alpha$ although both
appear very similar. Thus, the difference of observation results
at [N~$\rm{II}$]6583 and H$\alpha$ may be explained. When
carefully comparing the computed position-velocity diagrams of
[N~$\rm{II}$]6583 with our observations we do not find a certain
range of $\phi$ well constrained (possibly $\phi>$45$^\circ$).

According to the models of Hartigan et al. (1987), which are in
general made for H$\alpha$ profiles, when $\phi$=90$^\circ$ the
two peaks arise from different areas on the bow shock; the high
radial velocity component arises from near the apex, and the low
radial velocity from the wings. As $\phi$ decreases, the
distinction between peaks becomes less clear, since the expansion
of postshock material distributes emission over a range of radial
velocities. Toward Nebu.~Cw we have explicitly seen how the fast
velocity $V_1$ component and the slow $V_2$ one vary with
positions in [N~$\rm{II}$]6583 instead of in H$\alpha$. At
Y=--21.0$''$--~--15$''$, [N~$\rm{II}$]6583 profiles only show the
slow velocity $V_2$ component, which might be due to the wing of
the shock. At Y=--12.0$''$--~--9$''$, the profiles are double
peaked with both the slow and the fast velocity components ($V_2$
and $V_1$), which could be from the transition region of the wing
and the apex. At Y=--6$''$--0$''$, the profiles are brightest with
only the fast velocity $V_1$ emission that is from the shock apex.
At Y=3$''$--12$''$, the profiles present increasingly getting
weaker emission with broader $V_1$ component which is likely due
to the expansion of postshock material. Similar analyses of
velocity structures obtained with line profiles of
[Fe~$\rm{II}$]$\lambda$1.644$\mu$m lines were given by Pyo et al.
(2002) for L1551~IRS~5.

\subsection{Possible origin and more on the nature of the giant bow shocks}

As we have known from their prominent similarities of PNe, SNRs,
H$\rm{II}$ regions and HH objects, and therefore lots of confusion
made in history, now we are about to re-examine the nature of
these giant arcs. Firstly they are not PNe which are usually
symmetrical and possess small scales. Secondly, it's very unlikely
for them to be from SNRs. By far no SNRs have been found near
these arcs. Thirdly, can they belong to the extremely large
H$\rm{II}$ region, the Barnard's Loop (BL)? The much more diffuse
nebulae to the west (Fig.~1) are most likely due to BL. Heiles et
al. (2000) find that in BL $n_{\rm e}\sim$2.0~cm$^{-3}$, which is
one to two orders of magnitude lower than the values of the
nebulosities we have measured. If these arcs were related to BL,
it would be a real problem for so diffuse nebulae to be locally
enhanced into such well defined structures. So it becomes highly
possible that these arcs are HH shocks. In view of rather low
radial velocities ($V_{\rm rHELIO}\sim-$20--$-$40~km~s$^{-1}$, or
$V_{\rm rLSR}\sim$ --~5$-$--~25 km~s$^{-1}$), more extended
morphology features, and the narrow line width ($\Delta{\it
V}_{\rm FWZI} \sim$ 45.0~km~s$^{-1}$), the bow shocks observed by
us cannot be taken as normally defined HH objects. Additionally,
there is neither YSO nor molecular dense gas nearby these shocks,
these striking features could come from somewhere far away. The
spatial extent which HH objects or jets can flow through, has been
realized from typical 0.3 to 3 pc, and then even 10 pc, we now
have no reason to suspect that some HH objects could even pass
through a larger scale. These nebulosities are possibly the case.

As already intuitively suggested by Fig.~1 and kinematically
analyzed in previous sections, it turns out that these
nebulosities loose velocity and excitation, but gain size when
moving southwards, i.e. going south from Nebu.~K to C. So the
center of expansion is somewhere in the north. Inspecting the
large scale distribution of the giant bipolar HH flows
investigated by Reipurth et al. (1998) and Mader et al. (1999),
Nebu.~C is aligned with HH~404, HH~403 and HH~127 (Fig.~10), which
are on the direction with a position angle (P.A.) of 16$^\circ$,
while HH~131, HH~127 and the L1641-N VLA source are on a line with
P.A.$\sim$11$^\circ$. HH~404, HH~403 and HH~127 form a giant
bipolar HH flow driven by L1641-N VLA (Reipurth et al. 1998). We
wonder if Nebu.~C is associated with the giant HH flow or not. If
HH~131, Nebu.~C and L were driven by the VLA source, the distance
that the bow shocks have moved through would be $d\sim$15~pc, the
age of the shocks would be about 2$\times$10$^5$~yr provided the
average velocity of the shocks is $V\sim$100~km~s$^{-1}$ when they
flow through ISM. However, the VLA source is a deeply embedded
YSO, and therefore unlikely to be the source. Meanwhile we note
that the features A,B, C, and D are located close to the shocks
(Mader et al. 1999), the brightest two of which, B and D, have
been identified as HH 480 and 479, respectively (Yang \& Yao,
2000). It is interesting that B and D present unusual morphologies
too. Their positions are indicated in Fig.~10. They are likely to
be related with the nebulosities discussed in the paper. HH~479,
127 and 61/62 just emerge at the western edge of the Orion~A GMC,
while HH~131 and the nebulosities nearby are totally located out
of the GMC. The bright features (Nebu.~C/Cw, L and K) might have
been ejected out of the GMC by some source in M42 or NGC1999. And
later on they continue expanding in rather sparse ISM and keep
well their morphology. Proper motion studies are highly desirable
to constrain the location of their energy source.

\section{Conclusions}
We have carried out echelle spectrographic observations toward two
giant bow shocks (Nebu.~Cw and L) associated with HH 131 in Orion
using the HDS at the Subaru Telescope. With the low dispersion
spectrograph at the NAO 2.16~m telescope, spectra of Nebu.~C, L
and K have been obtained. Position-velocity distributions of
Nebu.~Cw and L have been analyzed from the long-slit spectra
observed with the HDS H$\alpha$ narrowband filter. We summarize
the conclusions as follows:

Strong [S~$\rm{II}$]$\lambda$$\lambda$ 6717/6731, H$\alpha$ and
[N~$\rm{II}$]6583 emission lines have been detected towards
Nebu.~C, Cw, L and K. The electron densities of all the observed
nebulosities are low ($n_{\rm e}\sim$10$^2$~cm$^{-3}$), which have
been deduced from the line ratio of [S~$\rm{II}$]6717/6731. The
excitation and electron density tend to decrease from K, L to C,
so there is a tendency for the flow to become less excited and
less ionized when going further to the south, where the most
extended (and presumably evolved) objects are seen.

Double kinematic signatures have been found in Nebu.~Cw from its
[N~$\rm{II}$]6583 profiles. The observed H$\alpha$ profiles of Cw
are almost symmetric. Bow shock models agree with the observed
position-velocity diagrams of [N~$\rm{II}$] to a higher degree
than with H$\alpha$ spectra, and a spatially resolved bow shock
with its wing, apex and postshock is likely revealed near Cw from
the [N~$\rm{II}$] emission.

The fastest radial velocity of Nebu.~Cw is
$V_r\sim$--18.0~km~s$^{-1}$. When the flow at L goes to the south,
it slows down. The fastest radial velocity of L is
--45.0~km~s$^{-1}$ and the slowest value is about
--18.3~km~s$^{-1}$. The radial velocity gradient is about
200~km~s$^{-1}$~pc$^{-1}$. The similarity of the fastest radial
velocity of Cw to the slowest value of L and their spatial
connection indicate that they are physically associated.

The high line ratio of
[S~$\rm{II}$]$\lambda\lambda$(6717+6731)/H$\alpha$ and the
non-detection of [O~$\rm{III}$]5007 suggest that Nebu. C, Cw, L
and K are characterized by low to intermediate excitation HH
spectra.

No certain source has been assigned to drive these giant features.
They might have been ejected out of the Orion~A GMC, later on they
continue expanding in rather sparse ISM and keep well their
morphology.

\begin{acknowledgements}

The authors wish to acknowledge the efforts and excellent support
of the SUBARU staff members during observations, and especially
thank Akito Tajitsu and Aoki Wako for their helpful discussions in
the data reductions. We also acknowledge the staff members of NAO
2.16~m telescope and BATC Beijing groups for their efforts and
helpful support during the observations of this project. This
research was supported by NSFC grants 10133020, 10243004, 10473022
and G19990754.
\end{acknowledgements}

\clearpage

\begin{deluxetable}{lccccll}
\tablecaption{Log of observations}

\tablehead{

Object & $\alpha$(J2000) & $\delta$(J2000)& Slit P.A.~($^\circ$) &
Slit Length($''$) &Exp.~T. (s)&Filter}

\startdata
\multicolumn{7}{l}{........spectrograph at NAO 2.16~m:} \\
Nebu. C & 05:33:51.36 & -08:42:38.7& 0   &240 &1800$\times$3 & \\
Nebu. L & 05:34:27.12 & -08:34:24.3& 0   &240 &1800$\times$1 & \\
Nebu. K & 05:34:19.37 & -08:32:01.7& 0   &240 &1800$\times$1 & \\
\tableline
\multicolumn{7}{l}{........HDS at SUBARU 8.2~m:} \\
Nebu. Cw& 05:33:48.98 & -08:42:45.6& 0   & 5 &1800$\times$1 & KV~408\\
Nebu. Cw& 05:33:49.04 & -08:42:46.5& 0   &60 &1800$\times$2 & H$\alpha$\\
Nebu. L & 05:34:26.27 & -08:34:18.0& 150 &60 &1800$\times$2 & H$\alpha$\\

\enddata

\end{deluxetable}

\clearpage

\begin{deluxetable}{lllcrcrcr}
\tablecaption{Emission line intensities with the NAO 2.16~m}

\tablehead{
&\multicolumn{2}{c}{Identification} && \multicolumn{5}{l}{Integrated Intensity\tablenotemark{a}}\\
\tableline &\multicolumn{2}{c}{} && Nebu.~C && Nebu.~L && Nebu.~K
}

\startdata

&  H$\beta$      &4861  &&  39\tablenotemark{b} &&   39\tablenotemark{b} && 71\tablenotemark{b}\\
& [O~$\rm{I}$]   &6300  && ...          &&   29\tablenotemark{b} &&...\\
& [N~$\rm{II}$]  &6548  &&  17\tablenotemark{b} &&   19\tablenotemark{b} && 16\tablenotemark{b}\\
&  H$\alpha$     &6563  && 100 &&  100 &&100\\
& [N~$\rm{II}$]  &6583  &&  62 &&   57 && 42\\
& [S~$\rm{II}$]  &6717  &&  95 &&   79 && 50\\
& [S~$\rm{II}$]  &6731  &&  64 &&   57 && 39\\

\tableline\\\tableline
&$n_{\rm e}$(cm$^{-3}$)\tablenotemark{c} &&&  $<$40 &&  40 && 400\\
\enddata

\tablenotetext{a}{Integrated intensities of H$\alpha$ of Nebu.~C,
L and K are 8.7$\times$10$^{-15}$, 4.4$\times$10$^{-15}$ and
4.3$\times$10$^{-15}$~erg~cm$^{-2}$~s$^{-1}$, respectively.}

\tablenotetext{b}{Uncertain}

\tablenotetext{c}{Electron densities according to the line ratios
of [S~$\rm{II}$]$\lambda$6717/$\lambda$6731 (Osterbrock 1989)}

\end{deluxetable}

\clearpage

\begin{deluxetable}{llclr@{$\pm$}lr}
\tabletypesize{\scriptsize}

\tablecaption{Emission lines  and Gaussian fitting parameters for
Nebu.~Cw}

\tablehead{
\multicolumn{1}{l}{Line} & Center &
\multicolumn{1}{c}{Core} & FWHM & \multicolumn{2}{l}{~~~~Flux} &
\multicolumn{1}{c}{V$_{\rm r}^{\rm c}$}\\
& (\AA) &
\multicolumn{1}{c}{(10$^{-16}$erg~cm$^{-2}$~s$^{-1}$~\AA$^{-1}$)}
& (\AA) & \multicolumn{2}{l}{(10$^{-16}$erg~cm$^{-2}$~s$^{-1}$)}&
(km~s$^{-1}$)} 

\startdata
H$\beta$
& 4861.058&     11.07&   0.401 &    4.72 &   1.13 &-16.17\\
& 4861.250&     12.67&   0.211 &    2.85 &   0.60 &-4.32\\
& 4861.724\tablenotemark{a(b)}&    9.59&   0.564 &    5.76 &   1.59 &24.93\\

[O~$\rm{I}$]6300
& 6299.996&    5.01&   0.234 &    1.25 &   0.21 &-14.67\\
& 6300.338&    1.78&   0.202 &    0.38 &   0.19 &1.62\\
& 6300.745\tablenotemark{a} &    233.70&   0.162 &   40.21 &   2.15 &21.00\\
& 6300.987\tablenotemark{b} &   3.43&   0.160 &    0.58 &   0.15 &32.52\\

[O~$\rm{I}$]6363
& 6363.393&    3.45&   0.311 &    1.14 &   0.32 &-18.06\\
& 6363.738&    2.41&   0.104 &    0.27 &   0.11 &-1.79\\
& 6364.214\tablenotemark{a}&     81.31&   0.152 &   13.12 &   0.43 &20.65\\
& 6364.581\tablenotemark{b}&    1.43&   0.406 &    0.62 &   0.42 &37.95\\

[N~$\rm{II}$]6548
& 6547.718&    5.39&   0.330 &    1.90 &   0.22 &-15.21\\
& 6548.049&    2.31&   0.323 &    0.79 &   0.21 &-0.05\\
& 6548.638\tablenotemark{b}&    1.93&   0.465 &    0.95 &   0.31 &26.94\\

H$\alpha$
& 6562.481&     17.15&   0.592 &   10.81 &   0.82 &-14.58\\
& 6562.690&     13.70&   0.189 &    2.76 &   0.26 &-5.03\\
& 6563.231\tablenotemark{a(b)}&     14.18&   0.529 &    7.99 &   0.73 &19.70\\

[N~$\rm{II}$]6583
& 6583.067&     16.48&   0.167 &    2.92 &   0.22 &-17.45\\
& 6583.304&    8.95&   0.292 &    2.78 &   0.38 &-6.65\\
& 6583.989\tablenotemark{b}&    5.33&   0.387 &    2.19 &   0.51 &24.56\\

[S~$\rm{II}$]6717
& 6716.097&     24.20&   0.290 &    7.47 &   0.40 &-15.32\\
& 6716.377&     12.21&   0.231 &    3.00 &   0.32 &-2.81\\
& 6717.044\tablenotemark{b}&    2.89&   0.477 &    1.47 &   0.66 &26.98\\

[S~$\rm{II}$]6731
& 6730.543&     17.57&   0.345 &    6.45 &   0.50 &-12.17\\
& 6730.833&    3.25&   0.340 &    1.18 &   0.49 &0.76\\
& 6731.412\tablenotemark{b}&    3.22&   0.374 &    1.28 &   0.54 &26.56\\

\enddata

\tablenotetext{a}{All emission most likely from the sky light}

\tablenotetext{b}{Emission possibly from the sky light or partly
from the local star forming region}

\tablenotetext{c}{$\lambda_0$ see in Table~4}

\end{deluxetable}

\clearpage

\begin{deluxetable}{llccl}
\tablecaption{Line peaks of Nebu.~Cw}

\tablehead{

\colhead{$\lambda$} & \colhead{$\lambda_0$} & \colhead{Line}&
\colhead{Peak}      &\multicolumn{1}{c}{V$_{r}$} \\
\colhead{(\AA)}     & \colhead{(\AA)}       & \colhead{} &
\colhead{10$^{-16}$erg~cm$^{-2}$~s$^{-1}$~\AA$^{-1}$)} &
\colhead{(km~s$^{-1}$)} }

\startdata

  4861.195&4861.32\tablenotemark{b}  &H$\beta$     &         21.14&    -7.71\\
 6299.967 &6300.304\tablenotemark{a} &[O~$\rm{I}$] &          4.58&   -16.05\\
 6363.415 &6363.776\tablenotemark{a} &[O~$\rm{I}$] &          4.23&   -17.02\\
 6547.718 &6548.05\tablenotemark{a}  &[N~$\rm{II}$]&          5.39&   -15.21\\
 6562.679 &6562.80\tablenotemark{a}  &H$\alpha$    &         25.75&    -5.53\\
 6583.067 &6583.45\tablenotemark{b}  &[N~$\rm{II}$]&         16.28&   -17.45\\
 6716.107 &6716.44\tablenotemark{a}  &[S~$\rm{II}$]&         22.41&   -14.87\\
 6730.543 &6730.816\tablenotemark{a} &[S~$\rm{II}$]&         17.57&   -12.17\\

\enddata

\tablenotetext{a}{$\lambda_0$ from the atomic line list by the
University of Kentucky (see
http://www.pa.uky.edu/$\sim$peter/atomic)}

\tablenotetext{b}{$\lambda_0$ from `Allen's Astrophysical
Quantities', 1999}

\end{deluxetable}

\clearpage

\begin{figure}
\includegraphics[width=15cm]{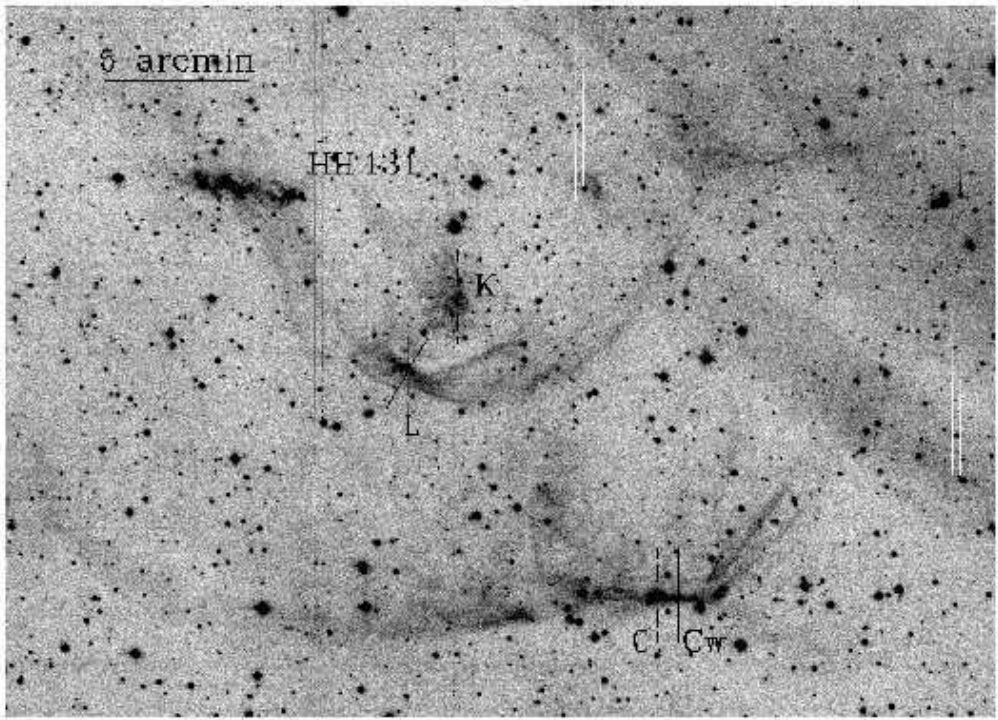}
\caption{Image of the nebulosities near HH~131 with exposure time
of 3600 sec, which was obtained with the [S~$\rm{II}$] narrow-band
filter ($\lambda$c=6725~\AA, $\Delta$$\lambda$=50~\AA) using the
0.6/0.9 cm Schmidt telescope at Xinglong Station of National
Astronomical Observatories, Chinese Academy of Sciences. Slit
positions at Nebu.~C, Cw, L and K in the low- and high-resolution
spectroscopic observations are indicated by dashed and solid
lines, respectively. East is to the left and north is up.}
\end{figure}
\clearpage

\newpage
\begin{figure}
\includegraphics[height=22cm,width=17cm]{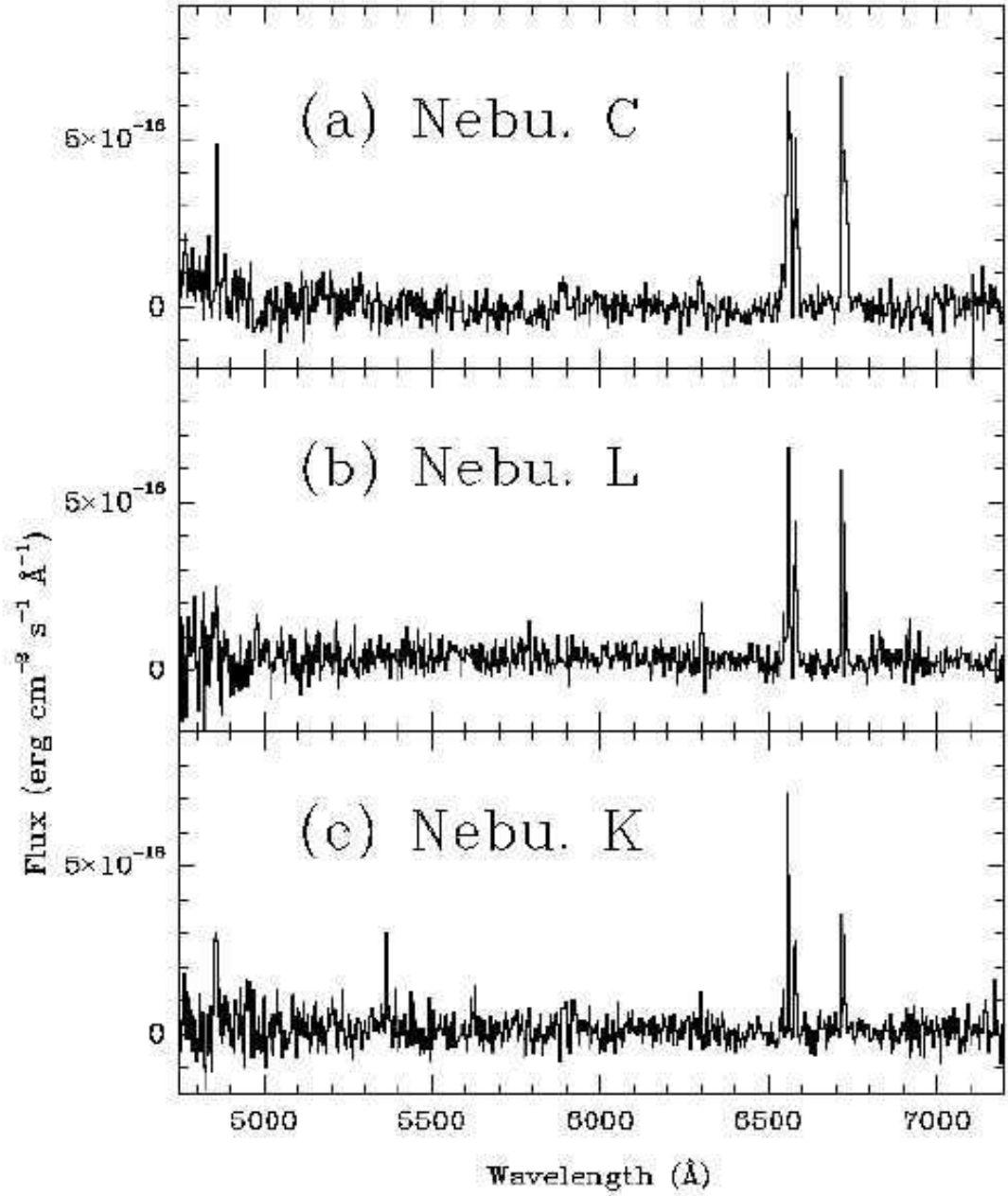}
\caption{Low resolution spectra of Nebu.~C, L and K with the NAO
2.16~m telescope.}
\end{figure}

\newpage
\begin{figure}
\includegraphics[height=22cm,width=17cm]{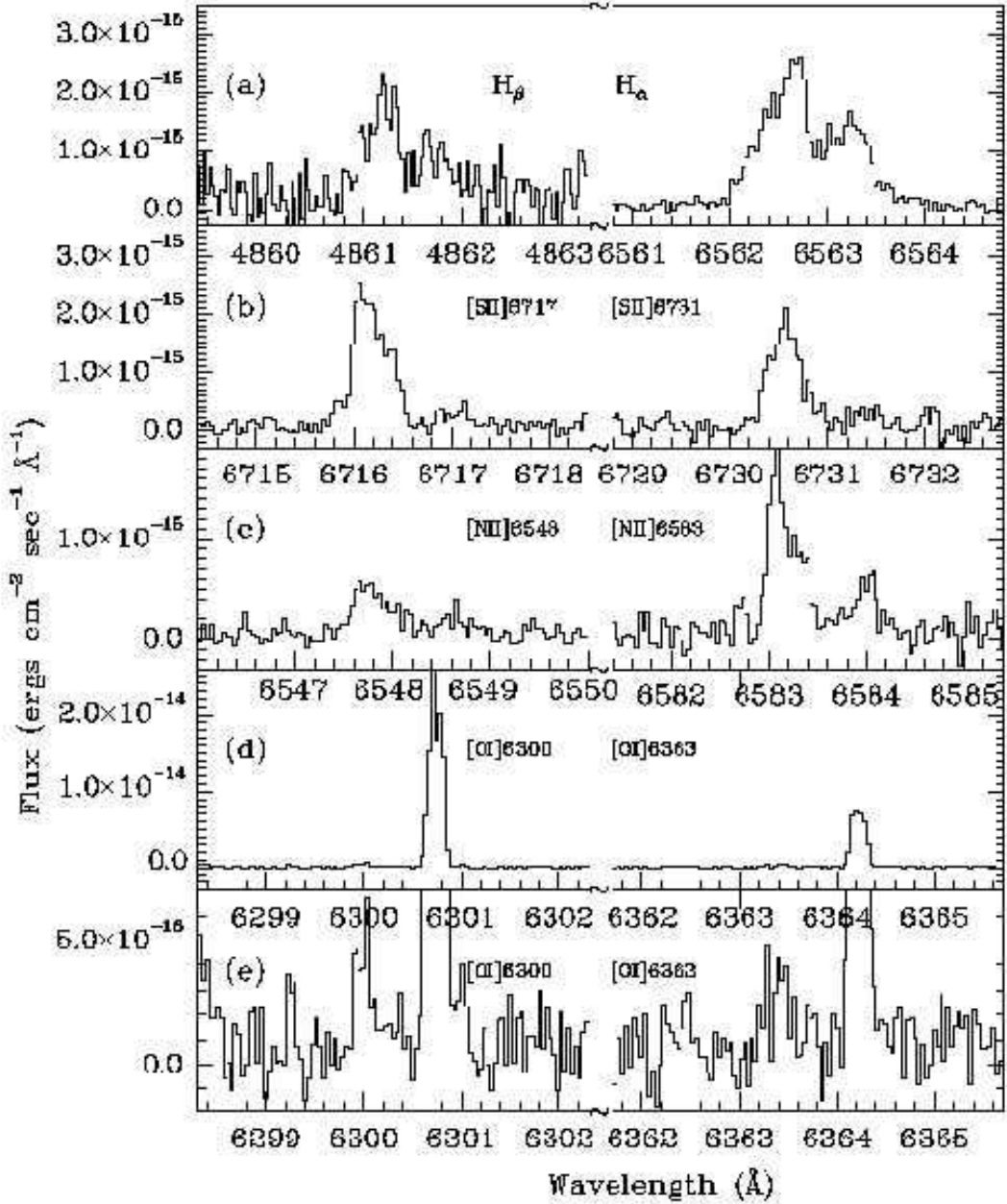}
\caption{HDS echelle line spectra of Nebu.~Cw obtained with the
filter StdYc or KV~408. Exposure time is 1800 sec.}
\end{figure}

\newpage
\begin{figure}
\includegraphics[width=17cm]{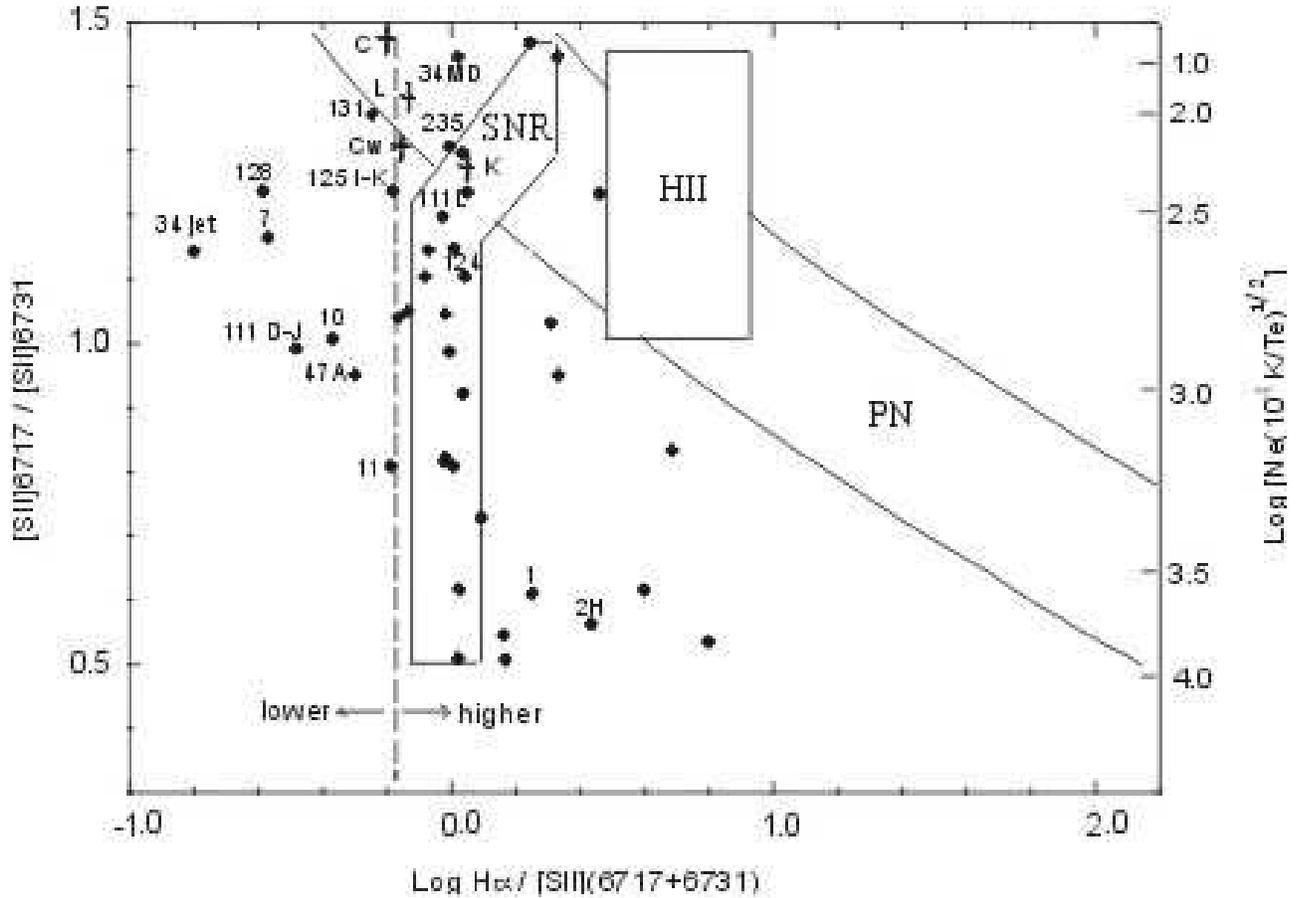}
\caption{Diagrams of log~H$\alpha$/[S~$\rm{II}$](6717+6731) versus
[S~$\rm{II}$]$\lambda$6717/$\lambda$6731 for various emission-line
objects. The positions of Nebu.~C, Cw, L and K are indicated with
crosses. HH objects are indicated with filled circles (data from
the compilation by Raga~et~al.~1996), some well known HH objects
are labeled. The regimes of PNe, SNRs and H~$\rm{II}$ regions
(adopted from the figures by Sabbadin~et~al.~1977 and Meaburn \&
White 1982) are marked. The dashed vertical line gives the
division between high/intermediate and low excitation of HH
objects.}
\end{figure}

\newpage
\begin{figure}
\includegraphics[width=15cm,height=12cm]{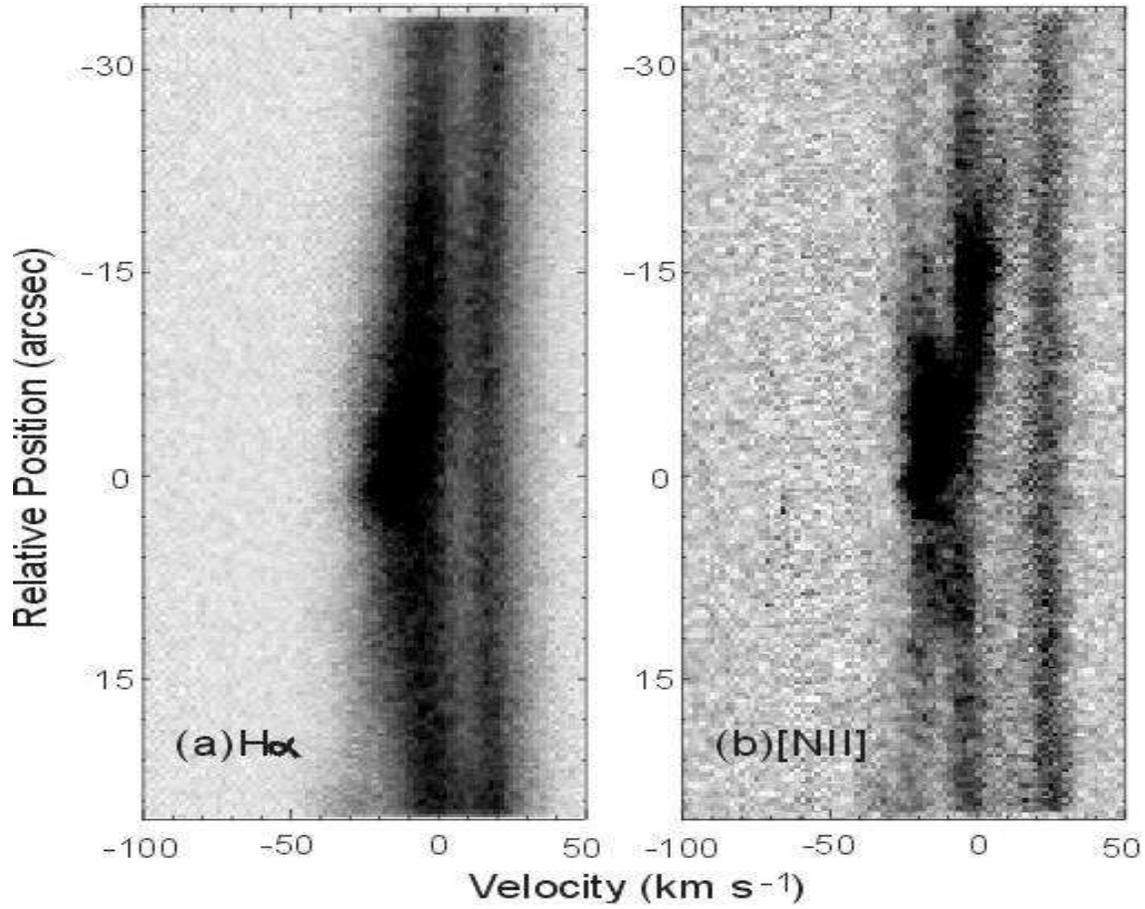}
\caption{The HDS long-slit gray-scaled position-velocity diagrams
for Nebu.~Cw  at (a) H$\alpha$; and (b)
[N~$\rm{II}$]$\lambda$6583~\AA. The vertical axis represents the
slit direction (see Fig.~1), and the horizontal follows the
wavelength dispersion.}
\end{figure}

\newpage
\begin{figure}
\includegraphics[width=15cm,height=12cm]{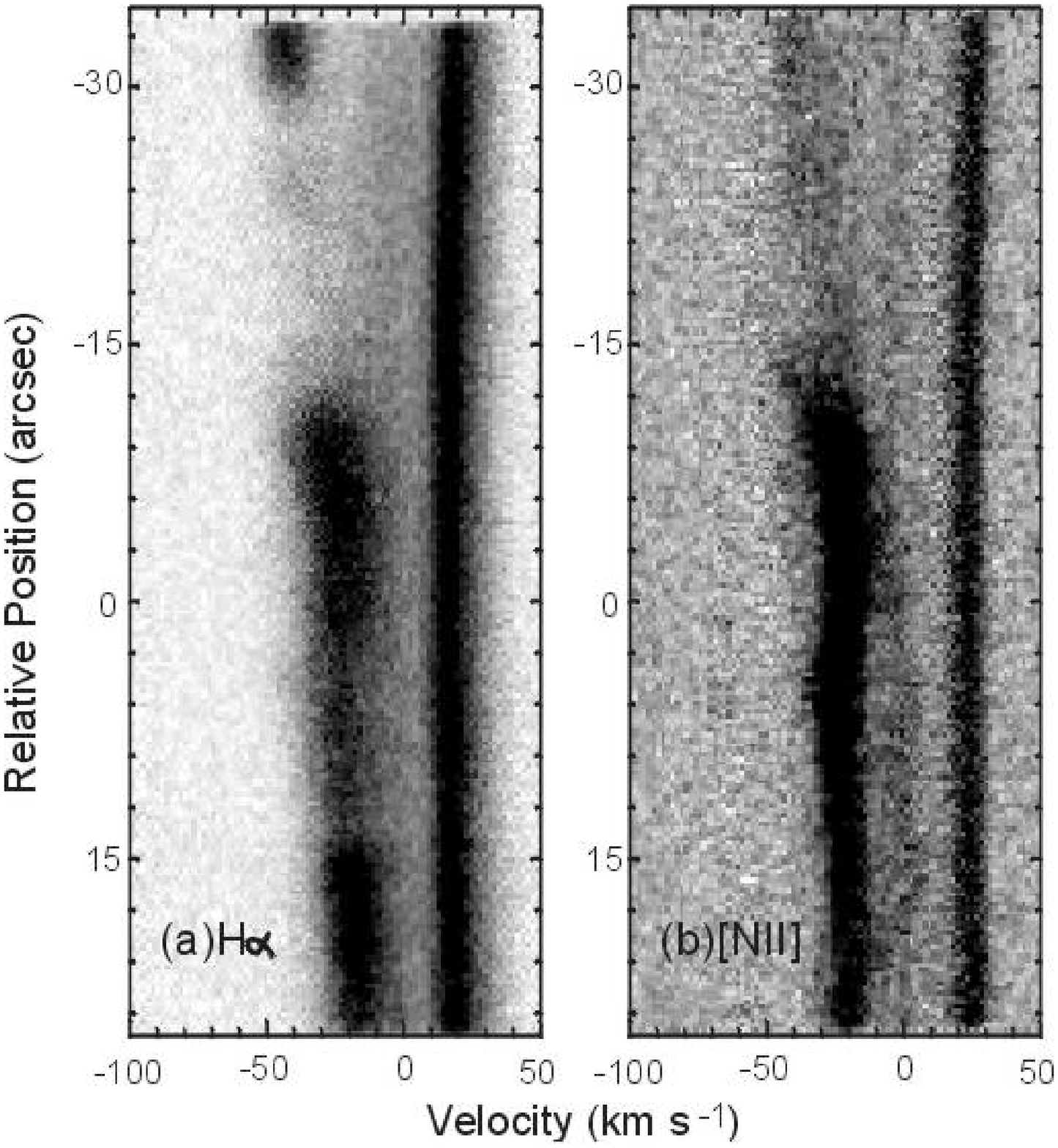}
\caption{Same as Fig.~5 but for Nebu.~L.}
\end{figure}

\newpage
\begin{figure}
\includegraphics[height=22cm,width=18cm]{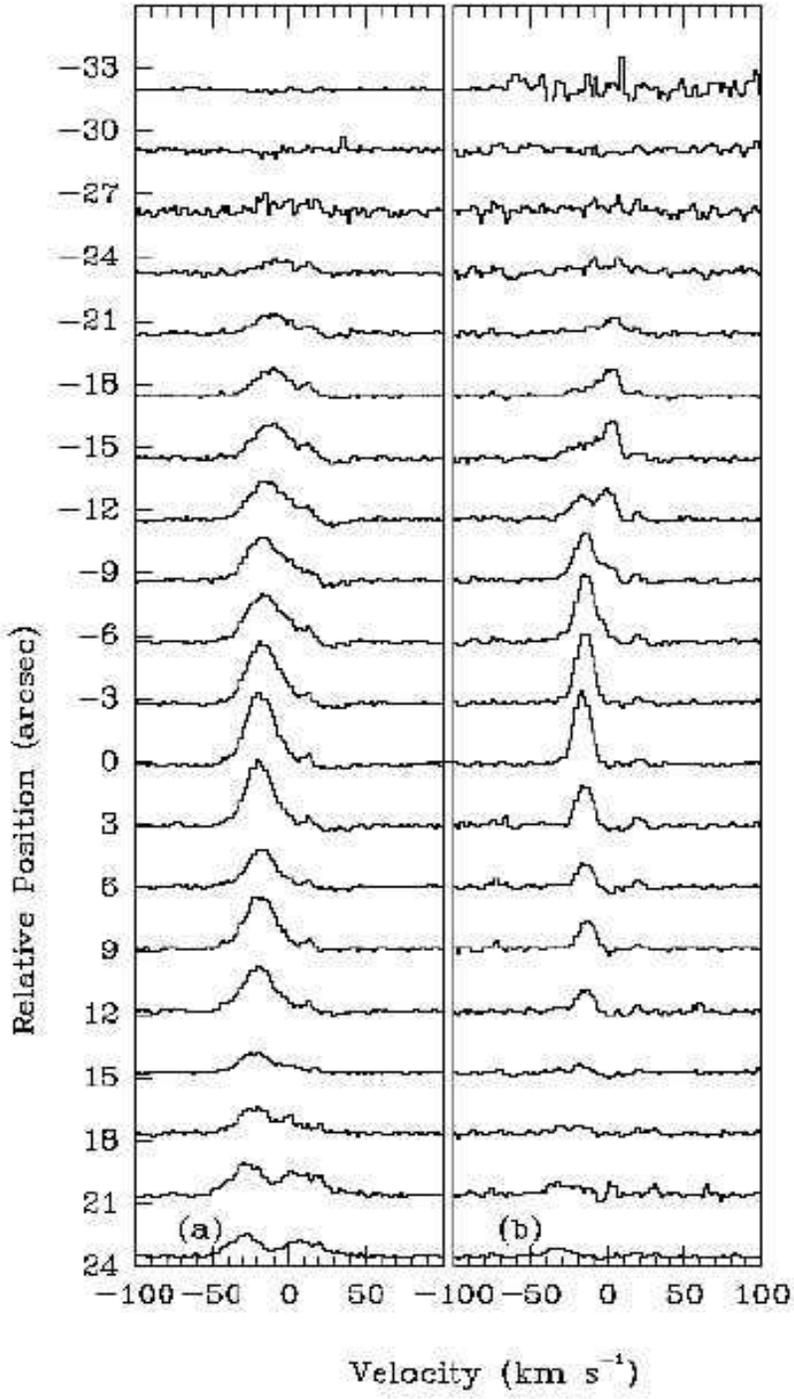}
\caption{Profiles of twenty positions of Nebu.~Cw at (a)
H$\alpha$; and (b) [N~$\rm{II}$]. Intensities in (b) are
multiplied by 2.}
\end{figure}

\newpage
\begin{figure}
\includegraphics[height=22cm,width=18cm]{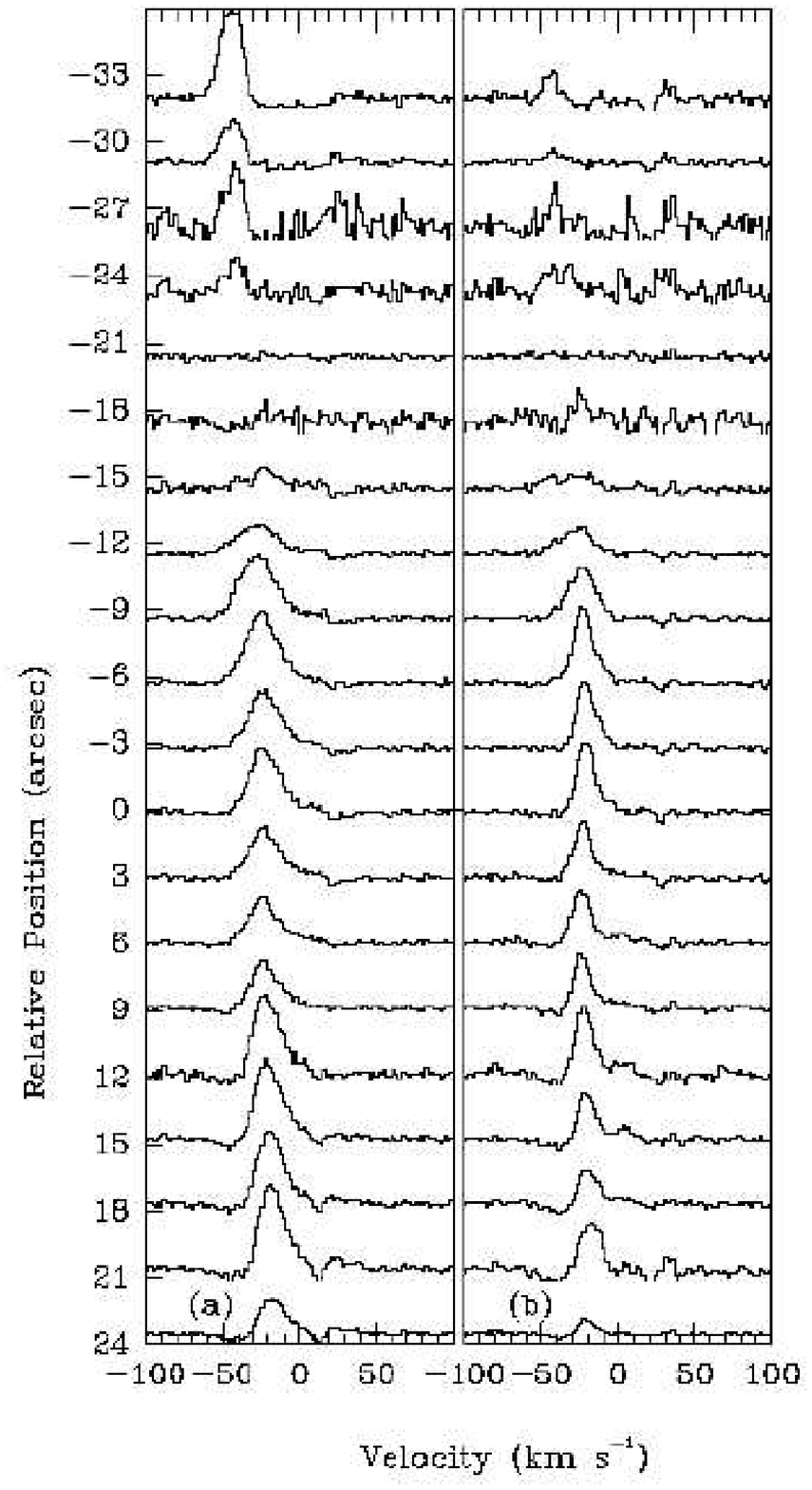}
\caption{Same as Fig.~7 but for Nebu.~L.}
\end{figure}

\newpage
\begin{figure}
\includegraphics[height=12cm,width=17cm]{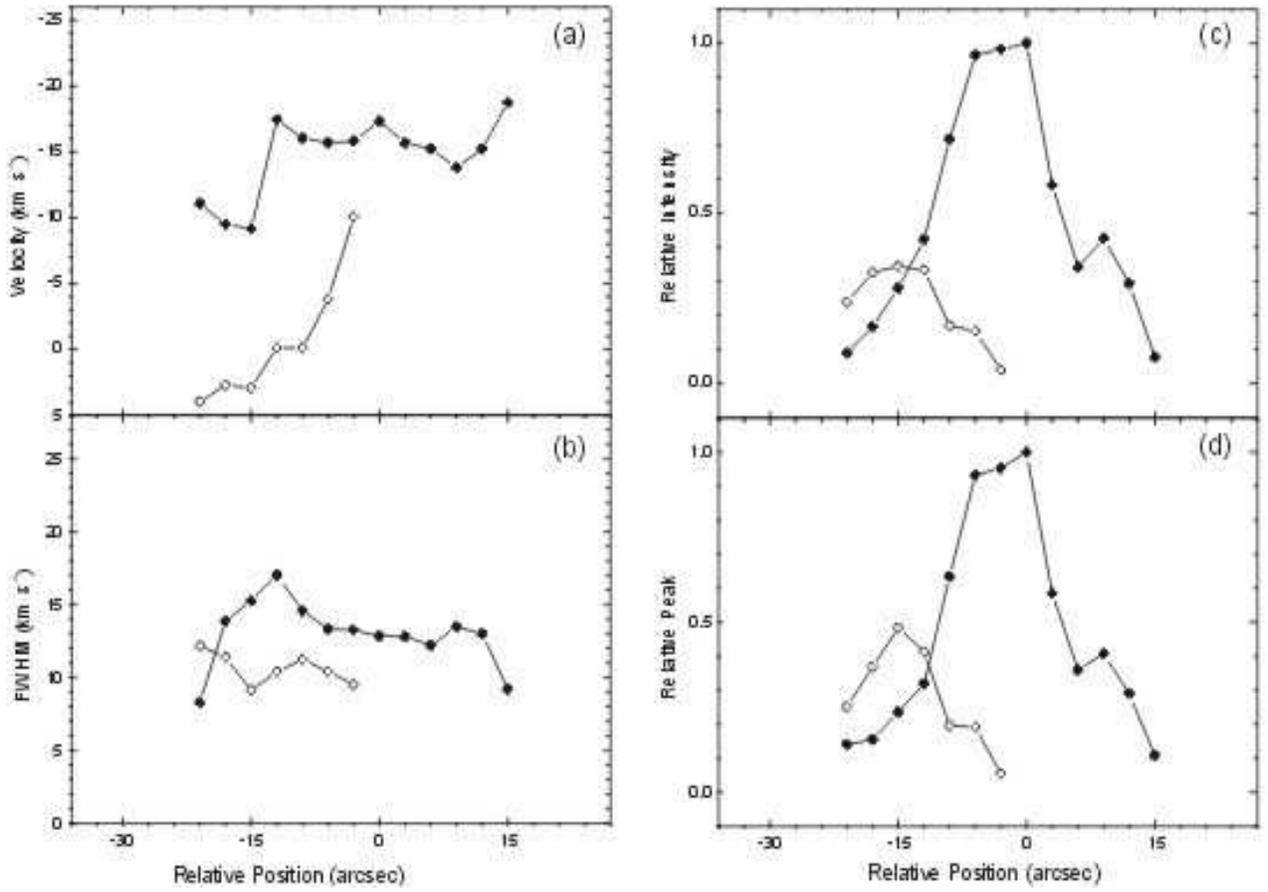}
\caption{Plots of the Gaussian fitting parameters of the
[N~$\rm{II}$]6583 profiles at various positions for Nebu.~Cw in
diagrams (a)--(d). Filled and open circles represent the V$_1$ and
V$_2$ components, respectively.}
\end{figure}

\newpage
\begin{figure}
\includegraphics[width=14cm]{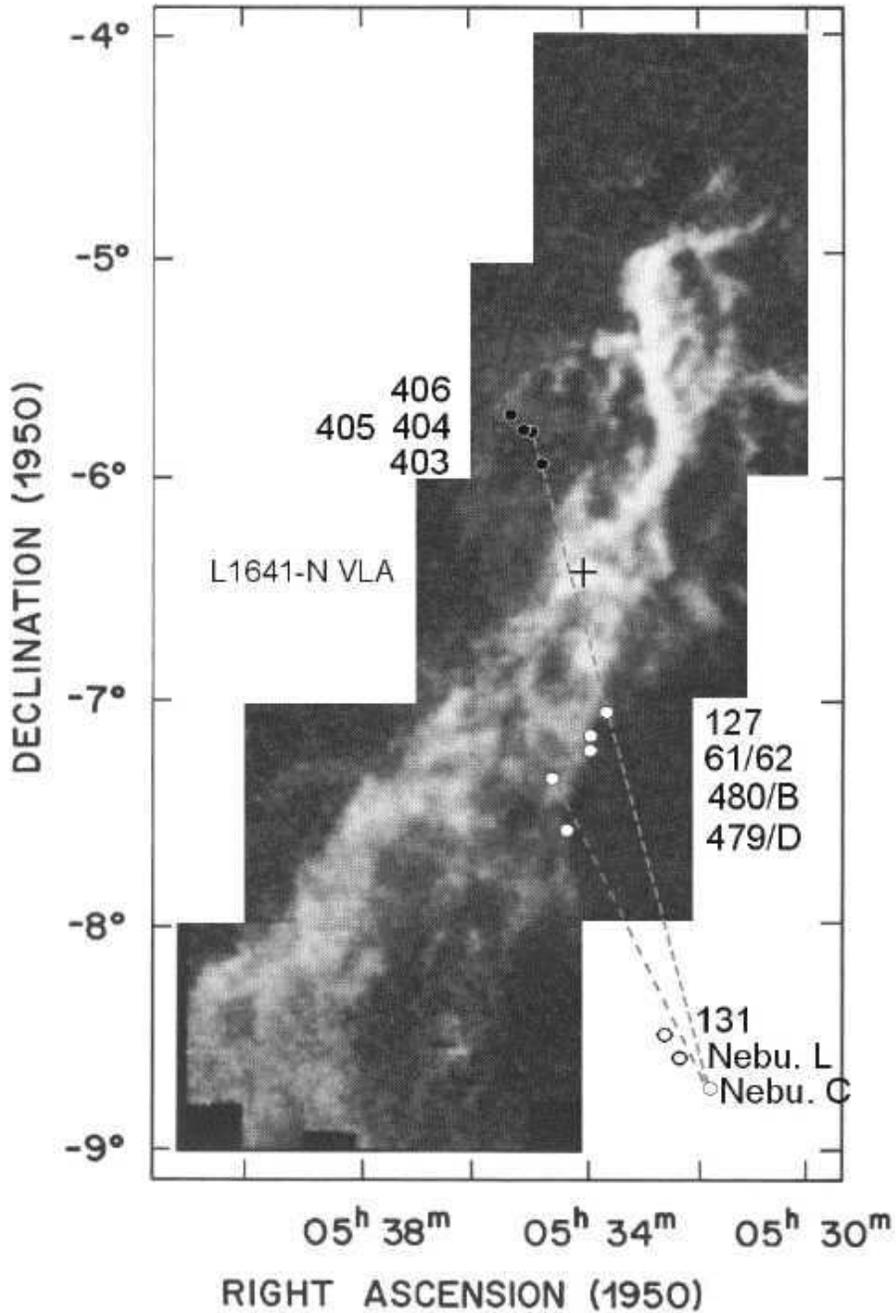}
\caption{Gray-scaled $^{13}$CO emission (adapted from Fig.~1 of
Bally et al. 1987) superposed with the positions of HH objects
discussed in the paper. The latter are indicated with the black
filled circles for the HH objects located to the north of the
L1641-N VLA source by the black cross, and the white filled
circles for the southern objects.}
\end{figure}

\end{document}